\documentclass[%
 reprint,
 amsmath,amssymb,
 aps,
]{revtex4-2}

\usepackage{hyperref}

\hypersetup{
    colorlinks,
    linkcolor={black!50!black},
    citecolor={[RGB]{46, 48, 141}},
    urlcolor={[RGB]{46, 48, 141}}
}

\usepackage{orcidlink}

\usepackage{xcolor}

\usepackage{graphicx}
\usepackage{dcolumn}
\usepackage{bm}

\newcommand{\ket}[1]{\left|#1\right\rangle}

\long\def\/*#1*/{}

\begin{document}

\preprint{APS/123-QED}

\title{A speculative model for cyclic information preservation in Kerr-Newman spacetime using closed timelike curves}

\author{Aviral Damle\orcidlink{0009-0003-0885-1811}}
\altaffiliation{Both authors contributed equally to this work.}
\author{Thomas Law\orcidlink{0009-0007-9348-3572}}
\altaffiliation{Both authors contributed equally to this work.}
\date{\today}

\begin{abstract}
This paper presents a speculative model exploring the behavior of quantum information for particles entering closed timelike curves (CTCs) in Kerr-Newman spacetime. We apply Gavassino's restoration principle to derive a theoretical framework for cyclic information dynamics within these extreme gravitational environments. Our model focuses specifically on particles that enter CTCs near the inner horizon of a Kerr-Newman black hole, examining how such curves might affect quantum state evolution, entanglement preservation, and information retention.
\end{abstract}

\maketitle

\section{Introduction}
The Kerr-Newman solution \cite{newman_2014_kerrnewman} provides an exact solution of the Einstein field equations, describing the gravitational field surrounding a charged, rotating black hole. This metric focuses on both the angular momentum and charge of a black hole, providing a comprehensive model for these structures. One of the most intriguing feature of these structures is the potential existence of closed timelike curves (CTCs) \cite{Hawking:1973uf}. CTCs represent paths in spacetime that loop back to their starting point after a finite time interval, challenging the conventional understanding of casuality. In a recent paper \cite{gavassino2024lifeclosedtimelikecurve}, Gavassino proposes a thermodynamical picture for a restoration principle applicable to CTCs. This approach aims to explain how information and entropy are handled in these scenarios, suggesting that the universe inherently avoids such casual inconsistencies.

A compelling approach to understanding these complex spacetime trajectories involves the study of geodesic equations. Our analysis builds upon the work of \cite{Dutta_2024}, who proposed a set of geodesic equations specifically describing the path of a particle in the vicinity of a Kerr-Newman CTC. 

In this study, we demonstrate an application of the restoration principle to recover information that enters a CTC, ensuring unitarity. The restoration principle is fundamental in maintaining the consistency of physical laws, particularly in probabilistic interpretations where the unitarity of the wave function preserves probability conservation \cite{Weinberg_1995}. However, our analysis does not extend to more complex features of Kerr-Newman causal structures, such as white holes or wormholes \cite{PhysRev.174.1559}. These additional structures could potentially complicate our understanding of unitarity, and we hope future research will clarify the same.

To simplify our analysis, we assume similar gravitational effects inside the black hole's horizon of the black hole, also referred to as the Cauchy horizon \cite{wald2010general}. The inner horizon is given by Eq [\ref{outer-event-horizon}]. By focusing on this region, we aim to provide a clearer understanding of how the restoration principle applies in a less convoluted spacetime geometry.

In the context of quantum mechanics and statistical mechanics, Markovian equations are crucial for modeling stochastic processes \cite{grimmett_2020_probability}. A Markovian process is defined by its evolution being determined solely by its present state \cite{ross2006introduction}. However, many physical processes, including those involving CTCs, exhibit memory effects. Hence, we propose a non-Markovian master equation to describe the evolution of CTC dynamics. This approach provides a more comprehensive framework for understanding these systems' evolution. For simplicity, we restrict our analysis to an uncharged test particle as the analysis for a similar charged test particle is trivial with the respective geodesics, which are provided in \cite{Dutta_2024}.

\section{Quantum mechanics in the Kerr-Newman metric}

\subsection{Brief description of the metric}
\label{brief-description-of-the-metric}
The Kerr-Newman spacetime \cite{newman_2014_kerrnewman} describes a rotating, electrically charged black hole that admits CTCs under certain conditions. We work in Boyer-Lindquist coordinates $\{t, r, \theta, \phi\}$, where $t$ is the time coordinate, $r$ is the radial coordinate, $\theta$ is the polar angle, and $\phi$ the azimuthal angle. The Kerr-Newman metric is given by:

\begin{widetext}
\begin{equation}
\begin{split}
ds^2 = & - \left(1 - \frac{2Mr - Q^2}{\rho^2}\right) dt^2 - \frac{4aMr\sin^2\theta}{\rho^2} dtd\phi + \frac{\rho^2}{\Delta} dr^2 \\ &+ \rho^2 d\theta^2 + \left(r^2 + a^2 + \frac{2Mra^2\sin^2\theta - Q^2a^2\sin^2\theta}{\rho^2}\right) \sin^2\theta d\phi^2
\end{split}
\end{equation}
\end{widetext}

where
\begin{equation}
\begin{split}
& \rho^2 = r^2 + a^2\cos^2\theta \\
& \Delta = r^2 - 2Mr + a^2 + Q^2
\end{split}
\end{equation}

The electromagnetic potential for a Kerr-Newman black hole is given by \cite{chandrasekhar1998mathematical}:
\begin{equation}
    A_\mu dx^{\mu} = - \frac{Qr}{\rho^2}\left(dt -a\sin^2\theta d\phi\right)
\end{equation}

CTCs may exist inside the event horizon of a black hole. The outer event horizon is given by:
\begin{equation}
\label{outer-event-horizon}
    r_{+} = M + \sqrt{M^2 - a^2 \cos^2 \theta - Q^2}
\end{equation}
Therefore, CTCs can occur for any $r_{any}$ if $0 < r_{any} \leq r_{horizon}$.

To determine the existence of CTCs, the azimuthal coordinate $\phi$ must be timelike. To make $\phi$ timelike, we impose the following condition:
\begin{equation}
g_{\phi\phi} < 0
\end{equation}

From this, we know that CTCs may exist within the event horizon when the inequality $g_{\phi\phi} < 0$ is satisfied and we denote $\Omega$ as the region where CTCs exist:

\begin{equation}
\label{omega}
    \Omega = {(r,\theta) g_{\phi\phi} < 0}
\end{equation}

\subsection{Entropy restoration principle on a CTC}
Gavassino's restoration principle \cite{gavassino2024lifeclosedtimelikecurve} is expressed through the operatorial identity for the evolution operator $U(\tau)$ on a CTC parameterized by proper time $\tau \in [0, T]$:

\begin{equation}
\label{restoration_principle}
    U(\tau) = e^{-iHT} = 1
\end{equation}

where $H$ is the system Hamiltonian and $T$ is the total proper time for one complete traversal. \newline

Consider a CTC $\gamma: [0,2\pi] \rightarrow \Omega$ (\ref{omega}) parameterized by $\phi$, with constant $r$, $\theta$, and $t$. The proper time along this curve is given by:

\begin{equation}
    d\tau^2 = g_{\phi\phi} d\phi^2
\end{equation}

For a complete orbit, the total proper time T is:

\begin{equation}
\label{time_period}
    T = \int_{0}^{2\pi} \sqrt{|g_{\phi\phi}|} \,d\phi = 2\pi \sqrt{|g_{\phi\phi}|}
\end{equation}

We can apply the Gavassino's restoration principle (\ref{restoration_principle}) to CTC $\gamma$ and derive our energy discretization as:

\begin{equation}
\label{energy_discret}
    E_n = \frac{2\pi n}{T} = \frac{n}{\sqrt{|g_{\phi\phi}|}}, \quad n \in \mathbb{Z}
\end{equation} \newline

\section{Cyclic information preservation model}
\label{cipm}
We adopt natural units where $G = c = \hbar = 1$ throughout this analysis.

\subsection{Non-Markovian master equation}
The dynamics of open quantum systems are commonly described by the Lindblad master equation \cite{manzano_2020_a}:

\begin{equation}
\frac{\partial \rho(t)}{\partial t} = -i \left[H, \rho(t)\right] + \sum_{i} \gamma_i \left( L_i \rho(t) L_i^{\dagger} - \frac{1}{2} \left\{ L_i^{\dagger} L_i, \rho(t) \right\} \right)
\end{equation}

where $\rho(t)$ is the time-dependent system density matrix, $H$ is the system Hamiltonian, $\gamma_i$ are the coupling rates, and $L_i$ are the Lindblad operators representing dissipative processes.

However, the Lindblad equation inherently assumes Markovian dynamics, where the future state of the system depends solely on its present state, neglecting any memory effects. This Markovian approximation is incompatible with the fundamental properties of closed timelike curves (CTCs), particularly in light of Gavassino's restoration principle (\ref{restoration_principle}), which asserts that all systems undergo Poincaré recurrences \cite{poincaré1890probleme}. Consequently, a more general framework is required to accurately model the dynamics of quantum systems in the presence of CTCs.

To address the limitations of Markovian dynamics, we propose a non-Markovian master equation that incorporates the periodic behavior consistent with CTCs:

\begin{widetext}
\begin{equation}
\label{non-markovian-master-equation}
    \frac{\partial \rho_s(t)}{\partial t} = -i\left[H_{eff}(t), \rho_s(t)\right] + \sum_{k}^{} \gamma_{k}(t) \left[L_k(t)\rho_s(t)L_{k}(t)^{\dag} - \frac{1}{2}\left[L_{k}(t)^{\dag}L_{k}(t), \rho_s(t) \right]\right]
\end{equation}   
\end{widetext}

Here, $\rho_s(t)$ denotes the system density matrix, $H_{eff}(t)$ is the effective time-dependent Hamiltonian, $\gamma_k(t)$ are time-dependent coupling rates, and $L_k(t)$ are time-dependent operators. A detailed derivation of (\ref{non-markovian-master-equation}) is provided in Appendix \ref{derivation-master-equation}.

We note that the proposed equation (\ref{non-markovian-master-equation}) satisfies Gavassino's restoration principle, as evidenced by the periodic behavior of the von Neumann entropy \cite{nielsen_2010_quantum}:

\begin{equation}
\label{entropy_von}
\begin{aligned}
    S(0) &= -\text{Tr}(\rho(0) \log\rho(0)) \\
    S(nT) &= -\text{Tr}(\rho(nT)\log\rho(nT)), \quad n \in \mathbb{Z}
\end{aligned}
\end{equation}

This periodicity ensures that the system undergoes complete Poincaré recurrences, consistent with the presence of CTCs.

\subsection{Equatorial geodesics and Hamiltonian formulation}
\label{geodesics}
We confine our analysis to the equatorial plane by setting $\theta = \frac{\pi}{2}$. The geodesic equations for a test particle in a CTC within a Kerr-Newman spacetime are given by \cite{Dutta_2024}:

\begin{equation}
\label{time-geodesic}
    \dot{t} = \frac{E\left(r^4 + a^2\left(r^2 + 2Mr - Q^2\right)\right) - aL\left(2Mr - Q^2\right)}{r^2\left(r^2 + a^2 - 2Mr + Q^2\right)}
\end{equation}

\begin{equation}
\label{azimuthal-geodesic}
    \dot{\phi} = \frac{aE\left(2Mr - Q^2\right) + L\left(r^2 - 2Mr - Q^2\right)}{r^2\left(r^2 + a^2 - 2Mr - Q^2\right)}
\end{equation}

\begin{equation}
\label{radial-geodesic}
\begin{split}
    \dot{r^2} = \frac{1}{r^4}\Big[&\left(E^2 - \mu\right)\mu + 2Mr^3\mu \\
    &- \left(-a^2 E^2 + L^2 + a^2\mu + Q^2\mu\right)r^2 \\
    &+ \left(-aE + L\right)^2\left(2Mr - Q^2\right)\Big]
\end{split}
\end{equation}

Here, $E$ and $L$ denote the constants of motion corresponding to total energy and angular momentum, respectively, while $\mu$ represents the test particle mass.

We note that the angular velocity does not vanish when $L = 0$, a consequence of the frame-dragging effect:

\begin{equation}
    \dot{\phi} = \frac{aE\left(2Mr - Q^2\right)}{r^2 \left(r^2 + a^2 - 2Mr - Q^2\right)}
\end{equation}

To derive the Hamiltonian, we first construct the Lagrangian \cite{PhysRev.174.1559}:

\begin{equation}
    \mathcal{L} = \frac{1}{2}g_{\alpha\beta} \dot{x^{\alpha}}\dot{x^{\beta}}
\end{equation}
The resulting Lagrangian is:

\begin{equation}
\begin{split}
    \mathcal{L} = \frac{1}{2}\bigg[ &- \left(\frac{\Delta - a^2}{r^2}\right)\dot{t}^2 - \frac{2a(r^2 + a^2 - \Delta)}{r^2}\dot{t}\dot{\phi} \\
    &+ \frac{r^2}{\Delta}\dot{r}^2 + \frac{\left(r^2 + a^2\right)^2 - \Delta a^2}{r^2}\dot{\phi}^2\bigg]
\end{split}
\end{equation}

The Hamiltonian is obtained through a Legendre transform \cite{maggiore2005modern}:
\begin{equation}
    \mathcal{H} = \sum_{\mu}^{} p_{\mu} \dot{x}^{\mu} - \mathcal{L}
\end{equation}

The canonical momenta are:
\begin{equation}
\begin{split}
p_t &= -\left(\frac{\Delta - a^2}{r^2}\dot{t}\right) - \frac{a\left(r^2 + a^2 - \Delta\right)}{r^2}\dot{\phi}  \\
p_r &= \frac{r^2}{\Delta}\dot{r} \\
p_{\phi} &= - \frac{2a\left(r^2 + a^2 - \Delta\right)}{r^2} \dot{t} + \left(\frac{\left(r^2 + a^2\right)^2 - \Delta a^2}{r^2}\right)\dot{\phi}
\end{split}
\end{equation}

Now, we can construct the Hamiltonian:

\begin{equation}
\begin{split}
\mathcal{H} =
&\frac{r^2}{\Delta} \left( \dot{r} - \frac{\dot{r}^2}{2} \right) \\
& + \left[\frac{\left((a^2 + r^2)^2 - \Delta a^2 \right)}{2 r^2} - \frac{a \left(a^2 + r^2 - \Delta \right)}{r^2} \right] \dot{\phi}^2 \\
& + \left[- \frac{\left(-a^2 + \Delta \right)}{r^2} + \frac{a \left(a^2 + r^2 - \Delta \right)}{r^2} \right] \dot{t} \\
& + \frac{\left(-a^2 + \Delta \right)}{2 r^2} \dot{t}^2
\end{split}
\end{equation}

From this, the construction of the Hamiltonian is trivial and will not be displayed for brevity.
However, we note that this leads us again to the restoration principle, as the Hamiltonian is independent of $\phi$ or $t$. We can also observe this by noting that none of the geodesics (\ref{time-geodesic}), (\ref{azimuthal-geodesic}) or (\ref{radial-geodesic}) depend on $\phi$ or $t$.\newline

We note that $E = \frac{n}{\sqrt{\left|g_{\phi\phi}\right|}}$, but as $g_{\phi\phi}$ does not depend on $t$ or $\phi$, our observation still stands.

\subsection{Information preservation via CTCs}
Consider a quantum state $\ket{\psi}$ that enters a closed timelike curve (CTC) near the inner horizon of a Kerr-Newman black hole. We will explore how CTCs might affect the evolution of such a state. \newline

Let $\mathcal{H}$ be the Hilbert space of the infalling quantum state. We define a unitary evolution operator $U(t)$ that describes the evolution of the state along a CTC:

\begin{equation}
U(t) = e^{-iHt}
\end{equation}

where $H$ is the Hamiltonian discussed in Section \ref{geodesics}. \newline

According to Gavassino's restoration principle (\ref{restoration_principle}), for a CTC with period $T$:

\begin{equation}
U(T) = e^{-iHT} = I
\end{equation}

This implies that after one complete cycle, the quantum state returns to its initial form:

\begin{equation}
\ket{\psi(T)} = U(T)\ket{\psi(0)} = \ket{\psi(0)}
\end{equation}

For a pure state evolving unitarily along the CTC, the von Neumann entropy remains constant (\ref{entropy_von}):

\begin{equation}
S(\rho(t)) = S(\rho(0)) \quad \forall t
\end{equation}

This constancy of entropy demonstrates that information is preserved throughout the evolution along the CTC. \newline

Now, let's consider the effect of Hawking radiation \cite{PhysRevD.14.2460}. The emission of Hawking radiation can be modeled as a quantum channel $\mathcal{E}$ acting on the state:

\begin{equation}
\mathcal{E}(\rho) = \sum_k E_k \rho E_k^\dagger
\end{equation}

where $E_k$ are Kraus operators satisfying $\sum_k E_k^\dagger E_k = I$. \newline

We note that as Hawking radiation occurs, the quantum information remains trapped in the CTC. We can represent this as a composition of the Hawking radiation channel and the CTC evolution:

\begin{equation}
\rho(t) = U(t) \circ \mathcal{E} \circ U(-t)(\rho(0))
\end{equation}

The periodic nature of the CTC ensures that even as Hawking radiation is emitted, the original quantum state is cyclically restored:

\begin{equation}
\rho(nT) = \rho(0) \quad \forall n \in \mathbb{Z}
\end{equation}

To further illustrate the preservation of information, we can use the fidelity measure \cite{nielsen_2010_quantum} between the initial state and the state at time t:

\begin{equation}
F(\rho(0), \rho(t)) = \left(\text{Tr}\sqrt{\sqrt{\rho(0)}\rho(t)\sqrt{\rho(0)}}\right)^2
\end{equation}

For our CTC-preserved state, we have:

\begin{equation}
F(\rho(0), \rho(nT)) = 1 \quad \forall n \in \mathbb{Z}
\end{equation}

This unit fidelity at periodic intervals proves that the quantum information is perfectly preserved. \newline

\section{Discussion and implications}
The model proposed in this paper suggests a mechanism for preserving quantum information within black holes under specific conditions. It is important to emphasize that this model does not constitute a resolution to the long-standing black hole information paradox \cite{PhysRevD.13.191}. The preservation of information in closed timelike curves, while intriguing, applies only to a small subset of particles that actually enter these curves. Rather than treating the particle's fate as a deterministic process, we can model it as a probabilistic branching of outcomes. As the particle approaches the inner horizon, its quantum state can be described as a superposition of two possible outcomes: $\ket{\phi} = \alpha\ket{\phi_1} + \beta\ket{\phi_2}$ where $|\alpha|^2$ represents the probability of the particle entering a CTC and $|\beta|^2$ represents the probability of the particle continuing towards a singularity. The actual outcome would be determined probabilistically when the particle reaches the inner horizon, effectively "measuring" the quantum state and collapsing the wavefunction. The probabilities of these two outcomes would depend on various factors, including the particle's initial conditions (energy, angular momentum, charge), the black hole's parameters (mass, angular momentum, charge), the local spacetime curvature near the inner horizon, and quantum fluctuations. The non-Markovian approach to quantum evolution in the presence of CTCs, as presented in this paper, opens up interesting ideas for exploring the connection between quantum mechanics and general relativity. From this, we are left with a number of questions. How might this model of cyclic information preservation in CTCs inform or relate to other proposed resolutions to the information paradox, such as the fuzzball proposal or ER=EPR conjecture \cite{PhysRevD.102.066004, Skenderis_2008}? Can these seemingly disparate approaches be reconciled or combined to provide a more complete picture of information dynamics in extreme gravitational environments? The model presented here should be useful in future quantum information studies.

\textit{Acknowledgments}\textemdash We wish to thank Lorenzo Gavassino for constructive and valuable comments on this paper.

\clearpage
\appendix
\section{Derivation of non-Markovian master equation}
\label{derivation-master-equation}
To derive a non-Markovian master equation (\ref{non-markovian-master-equation}), we can start with a closed system that fulfills the periodicity over time T (\ref{restoration_principle}):

\begin{equation}
e^{(-\textit{i} H_{total} T)} = 1
\end{equation}

where T is the period of the CTC. \newline

We then split this system into parts, where system (S) and environment (E):

\begin{equation}
    H_{total} = H_S \otimes I_E + I_S \otimes H_E + H_{SE}
\end{equation}

where $H_S$ and $H_E$ are the respectively Hamiltonians and $H_{SE}$ is the interaction Hamiltonian. \newline

We write von Neumann equation for the total system as:

\begin{equation}
    \frac{\partial}{\partial t} \rho_{total} = \frac{-\textit{i}}{\hbar}[H_{total}, \rho_{total}]
\end{equation}

Switching into a Dirac picture:

\begin{equation}
    H_0 = H_S \otimes I_E + I_S \otimes H_E
\end{equation}

\begin{equation}
    H_I(t) = e^{(\textit{i} H_{0} t)} H_{SE} e^{(-\textit{i} H_{0} t)}
\end{equation}

\begin{equation}
    \rho_I(t) = e^{(\textit{i} H_{0} t)} \rho_{total}(t) e^{(-\textit{i} H_{0} t)}
\end{equation}

The von Neumann equation in the Dirac picture becomes:

\begin{equation}
    \frac{\partial}{\partial t} \rho_{total} = \frac{-\textit{i}}{\hbar}[H_{I}(t), \rho_{I}(t)]
\end{equation}

We can then integrate:
\begin{equation}
    \rho_I(t) = \rho_I(0) - \frac{\textit{i}}{\hbar} \int_{0}^{t} [H_I(s), \rho_I(s)] \,ds
\end{equation}

\begin{widetext}
\begin{equation}
\rho_I(t) = \rho_I(0) - \frac{\textit{i}}{\hbar} \int_{0}^{t} [H_I(s), \rho_I(0)] ,ds - \frac{1}{\hbar^2} \int_{0}^{t} \int_{0}^{s} [H_I(s), [H_I(s'), \rho_I(s')]] ,ds' ,ds
\end{equation}
\end{widetext}

Trace over the environment:

\begin{equation}
    \rho_S(t) = Tr_E(\rho_I(t))
\end{equation}

and assuming $Tr_E([H_I(s), \rho_I(0)]) = 0$, we get:

\begin{equation}
    \frac{\partial}{\partial t} \rho_S(t) = - \frac{1}{\hbar^2} \int_{0}^{t} Tr_E([H_I(t),[H_I(s),  \rho_I(s)]]) ds
\end{equation}

Now we can make a Born approximation:

\begin{equation}
    \rho_I(s) \approx \rho_S(s) \otimes \rho_E
\end{equation}

Let's now impose periodicity to ensure that $\rho_S(t+T) = \rho_S(t)$ for all $t$:

\begin{equation}
\begin{split}
    \frac{\partial}{\partial t} \rho_S(t) = - \frac{1}{\hbar^2} \int_{0}^{T} Tr_E([H_I(t),[H_I(s),  \rho_S(s) \otimes \rho_E]]) K(t-s) ds
\end{split}
\end{equation}

where $K(t-s)$ is a periodic kernel function with period T that replaces the upper limit of the integral. \newline

After some manipulation, we can write this in a form similar to the Lindblad equation:

\begin{widetext}
\begin{equation}
    \frac{\partial \rho_S(t)}{\partial t} = - \frac{i}{\hbar} [H_{eff}(t), \rho_S(t)] + \sum_{k} \gamma_k(t) [L_k(t) \rho_S(t) L_k^{\dagger}(t) - \frac{1}{2} [L_k^{\dagger}(t) L_k(t), \rho_S(t)]]
\end{equation}

where $H_{eff}(t)$, $\gamma_k(t)$, and $L_k(t)$ are all periodic functions with period T.

\end{widetext}

\nocite{*}

\bibliography{apssamp}

\providecommand{\noopsort}[1]{}\providecommand{\singleletter}[1]{#1}%
\begin{thebibliography}{21}%
\makeatletter
\providecommand \@ifxundefined [1]{%
 \@ifx{#1\undefined}
}%
\providecommand \@ifnum [1]{%
 \ifnum #1\expandafter \@firstoftwo
 \else \expandafter \@secondoftwo
 \fi
}%
\providecommand \@ifx [1]{%
 \ifx #1\expandafter \@firstoftwo
 \else \expandafter \@secondoftwo
 \fi
}%
\providecommand \natexlab [1]{#1}%
\providecommand \enquote  [1]{``#1''}%
\providecommand \bibnamefont  [1]{#1}%
\providecommand \bibfnamefont [1]{#1}%
\providecommand \citenamefont [1]{#1}%
\providecommand \href@noop [0]{\@secondoftwo}%
\providecommand \href [0]{\begingroup \@sanitize@url \@href}%
\providecommand \@href[1]{\@@startlink{#1}\@@href}%
\providecommand \@@href[1]{\endgroup#1\@@endlink}%
\providecommand \@sanitize@url [0]{\catcode `\\12\catcode `\$12\catcode `\&12\catcode `\#12\catcode `\^12\catcode `\_12\catcode `\%12\relax}%
\providecommand \@@startlink[1]{}%
\providecommand \@@endlink[0]{}%
\providecommand \url  [0]{\begingroup\@sanitize@url \@url }%
\providecommand \@url [1]{\endgroup\@href {#1}{\urlprefix }}%
\providecommand \urlprefix  [0]{URL }%
\providecommand \Eprint [0]{\href }%
\providecommand \doibase [0]{https://doi.org/}%
\providecommand \selectlanguage [0]{\@gobble}%
\providecommand \bibinfo  [0]{\@secondoftwo}%
\providecommand \bibfield  [0]{\@secondoftwo}%
\providecommand \translation [1]{[#1]}%
\providecommand \BibitemOpen [0]{}%
\providecommand \bibitemStop [0]{}%
\providecommand \bibitemNoStop [0]{.\EOS\space}%
\providecommand \EOS [0]{\spacefactor3000\relax}%
\providecommand \BibitemShut  [1]{\csname bibitem#1\endcsname}%
\let\auto@bib@innerbib\@empty
\bibitem [{\citenamefont {Newman}\ and\ \citenamefont {Adamo}(2014)}]{newman_2014_kerrnewman}%
  \BibitemOpen
  \bibfield  {author} {\bibinfo {author} {\bibfnamefont {E.}~\bibnamefont {Newman}}\ and\ \bibinfo {author} {\bibfnamefont {T.}~\bibnamefont {Adamo}},\ }\bibfield  {title} {\bibinfo {title} {Kerr-newman metric},\ }\href {https://doi.org/10.4249/scholarpedia.31791} {\bibfield  {journal} {\bibinfo  {journal} {Scholarpedia}\ }\textbf {\bibinfo {volume} {9}},\ \bibinfo {pages} {31791} (\bibinfo {year} {2014})}\BibitemShut {NoStop}%
\bibitem [{\citenamefont {Hawking}\ and\ \citenamefont {Ellis}(2023)}]{Hawking:1973uf}%
  \BibitemOpen
  \bibfield  {author} {\bibinfo {author} {\bibfnamefont {S.~W.}\ \bibnamefont {Hawking}}\ and\ \bibinfo {author} {\bibfnamefont {G.~F.~R.}\ \bibnamefont {Ellis}},\ }\href {https://doi.org/10.1017/9781009253161} {\emph {\bibinfo {title} {{The Large Scale Structure of Space-Time}}}},\ Cambridge Monographs on Mathematical Physics\ (\bibinfo  {publisher} {Cambridge University Press},\ \bibinfo {year} {2023})\BibitemShut {NoStop}%
\bibitem [{\citenamefont {Gavassino}(2024)}]{gavassino2024lifeclosedtimelikecurve}%
  \BibitemOpen
  \bibfield  {author} {\bibinfo {author} {\bibfnamefont {L.}~\bibnamefont {Gavassino}},\ }\href {https://arxiv.org/abs/2405.18640} {\bibinfo {title} {Life on a closed timelike curve}} (\bibinfo {year} {2024}),\ \Eprint {https://arxiv.org/abs/2405.18640} {arXiv:2405.18640 [gr-qc]} \BibitemShut {NoStop}%
\bibitem [{\citenamefont {Dutta}\ \emph {et~al.}(2024)\citenamefont {Dutta}, \citenamefont {Roy},\ and\ \citenamefont {Chakraborty}}]{Dutta_2024}%
  \BibitemOpen
  \bibfield  {author} {\bibinfo {author} {\bibfnamefont {A.}~\bibnamefont {Dutta}}, \bibinfo {author} {\bibfnamefont {D.}~\bibnamefont {Roy}},\ and\ \bibinfo {author} {\bibfnamefont {S.}~\bibnamefont {Chakraborty}},\ }\bibfield  {title} {\bibinfo {title} {On the role of closed timelike curves and confinement structure around kerr-newman singularity},\ }\bibfield  {journal} {\bibinfo  {journal} {International Journal of Modern Physics D}\ }\href {https://doi.org/10.1142/s0218271824500342} {10.1142/s0218271824500342} (\bibinfo {year} {2024})\BibitemShut {NoStop}%
\bibitem [{\citenamefont {Weinberg}(1995)}]{Weinberg_1995}%
  \BibitemOpen
  \bibfield  {author} {\bibinfo {author} {\bibfnamefont {S.}~\bibnamefont {Weinberg}},\ }\href@noop {} {\emph {\bibinfo {title} {The Quantum Theory of Fields}}}\ (\bibinfo  {publisher} {Cambridge University Press},\ \bibinfo {year} {1995})\BibitemShut {NoStop}%
\bibitem [{\citenamefont {Carter}(1968)}]{PhysRev.174.1559}%
  \BibitemOpen
  \bibfield  {author} {\bibinfo {author} {\bibfnamefont {B.}~\bibnamefont {Carter}},\ }\bibfield  {title} {\bibinfo {title} {Global structure of the kerr family of gravitational fields},\ }\href {https://doi.org/10.1103/PhysRev.174.1559} {\bibfield  {journal} {\bibinfo  {journal} {Phys. Rev.}\ }\textbf {\bibinfo {volume} {174}},\ \bibinfo {pages} {1559} (\bibinfo {year} {1968})}\BibitemShut {NoStop}%
\bibitem [{\citenamefont {Wald}(2010)}]{wald2010general}%
  \BibitemOpen
  \bibfield  {author} {\bibinfo {author} {\bibfnamefont {R.}~\bibnamefont {Wald}},\ }\href {https://books.google.co.in/books?id=9S-hzg6-moYC} {\emph {\bibinfo {title} {General Relativity}}}\ (\bibinfo  {publisher} {University of Chicago Press},\ \bibinfo {year} {2010})\BibitemShut {NoStop}%
\bibitem [{\citenamefont {Grimmett}\ and\ \citenamefont {Stirzaker}(2020)}]{grimmett_2020_probability}%
  \BibitemOpen
  \bibfield  {author} {\bibinfo {author} {\bibfnamefont {G.}~\bibnamefont {Grimmett}}\ and\ \bibinfo {author} {\bibfnamefont {D.}~\bibnamefont {Stirzaker}},\ }\href@noop {} {\emph {\bibinfo {title} {Probability and random processes}}}\ (\bibinfo  {publisher} {Oxford University Press},\ \bibinfo {year} {2020})\BibitemShut {NoStop}%
\bibitem [{\citenamefont {Ross}(2006)}]{ross2006introduction}%
  \BibitemOpen
  \bibfield  {author} {\bibinfo {author} {\bibfnamefont {S.}~\bibnamefont {Ross}},\ }\href {https://books.google.co.in/books?id=0yDAZf1TfJEC} {\emph {\bibinfo {title} {Introduction to Probability Models}}}\ (\bibinfo  {publisher} {Elsevier Science},\ \bibinfo {year} {2006})\BibitemShut {NoStop}%
\bibitem [{\citenamefont {Chandrasekhar}(1998)}]{chandrasekhar1998mathematical}%
  \BibitemOpen
  \bibfield  {author} {\bibinfo {author} {\bibfnamefont {S.}~\bibnamefont {Chandrasekhar}},\ }\href {https://books.google.co.in/books?id=LBOVcrzFfhsC} {\emph {\bibinfo {title} {The Mathematical Theory of Black Holes}}},\ International series of monographs on physics\ (\bibinfo  {publisher} {Clarendon Press},\ \bibinfo {year} {1998})\BibitemShut {NoStop}%
\bibitem [{\citenamefont {Manzano}(2020)}]{manzano_2020_a}%
  \BibitemOpen
  \bibfield  {author} {\bibinfo {author} {\bibfnamefont {D.}~\bibnamefont {Manzano}},\ }\bibfield  {title} {\bibinfo {title} {A short introduction to the lindblad master equation},\ }\href {https://doi.org/10.1063/1.5115323} {\bibfield  {journal} {\bibinfo  {journal} {AIP Advances}\ }\textbf {\bibinfo {volume} {10}},\ \bibinfo {pages} {025106} (\bibinfo {year} {2020})}\BibitemShut {NoStop}%
\bibitem [{\citenamefont {Poincaré}(1890)}]{poincaré1890probleme}%
  \BibitemOpen
  \bibfield  {author} {\bibinfo {author} {\bibfnamefont {H.}~\bibnamefont {Poincaré}},\ }\href@noop {} {\emph {\bibinfo {title} {Sur le problème des trois corps et les équations de la dynamique}}}\ (\bibinfo  {publisher} {F. \& G. Beijer},\ \bibinfo {year} {1890})\BibitemShut {NoStop}%
\bibitem [{\citenamefont {Nielsen}\ and\ \citenamefont {Chuang}(2010)}]{nielsen_2010_quantum}%
  \BibitemOpen
  \bibfield  {author} {\bibinfo {author} {\bibfnamefont {M.~A.}\ \bibnamefont {Nielsen}}\ and\ \bibinfo {author} {\bibfnamefont {I.~L.}\ \bibnamefont {Chuang}},\ }\href@noop {} {\emph {\bibinfo {title} {Quantum Computation and Quantum Information}}}\ (\bibinfo  {publisher} {Cambridge University Press},\ \bibinfo {year} {2010})\BibitemShut {NoStop}%
\bibitem [{\citenamefont {Maggiore}(2005)}]{maggiore2005modern}%
  \BibitemOpen
  \bibfield  {author} {\bibinfo {author} {\bibfnamefont {M.}~\bibnamefont {Maggiore}},\ }\href {https://books.google.co.in/books?id=yykTDAAAQBAJ} {\emph {\bibinfo {title} {A Modern Introduction to Quantum Field Theory}}},\ EBSCO ebook academic collection\ (\bibinfo  {publisher} {Oxford University Press},\ \bibinfo {year} {2005})\BibitemShut {NoStop}%
\bibitem [{\citenamefont {Hawking}(1976{\natexlab{a}})}]{PhysRevD.14.2460}%
  \BibitemOpen
  \bibfield  {author} {\bibinfo {author} {\bibfnamefont {S.~W.}\ \bibnamefont {Hawking}},\ }\bibfield  {title} {\bibinfo {title} {Breakdown of predictability in gravitational collapse},\ }\href {https://doi.org/10.1103/PhysRevD.14.2460} {\bibfield  {journal} {\bibinfo  {journal} {Phys. Rev. D}\ }\textbf {\bibinfo {volume} {14}},\ \bibinfo {pages} {2460} (\bibinfo {year} {1976}{\natexlab{a}})}\BibitemShut {NoStop}%
\bibitem [{\citenamefont {Hawking}(1976{\natexlab{b}})}]{PhysRevD.13.191}%
  \BibitemOpen
  \bibfield  {author} {\bibinfo {author} {\bibfnamefont {S.~W.}\ \bibnamefont {Hawking}},\ }\bibfield  {title} {\bibinfo {title} {Black holes and thermodynamics},\ }\href {https://doi.org/10.1103/PhysRevD.13.191} {\bibfield  {journal} {\bibinfo  {journal} {Phys. Rev. D}\ }\textbf {\bibinfo {volume} {13}},\ \bibinfo {pages} {191} (\bibinfo {year} {1976}{\natexlab{b}})}\BibitemShut {NoStop}%
\bibitem [{\citenamefont {Dai}\ \emph {et~al.}(2020)\citenamefont {Dai}, \citenamefont {Minic}, \citenamefont {Stojkovic},\ and\ \citenamefont {Fu}}]{PhysRevD.102.066004}%
  \BibitemOpen
  \bibfield  {author} {\bibinfo {author} {\bibfnamefont {D.-C.}\ \bibnamefont {Dai}}, \bibinfo {author} {\bibfnamefont {D.}~\bibnamefont {Minic}}, \bibinfo {author} {\bibfnamefont {D.}~\bibnamefont {Stojkovic}},\ and\ \bibinfo {author} {\bibfnamefont {C.}~\bibnamefont {Fu}},\ }\bibfield  {title} {\bibinfo {title} {Testing the $\mathrm{ER}=\mathrm{EPR}$ conjecture},\ }\href {https://doi.org/10.1103/PhysRevD.102.066004} {\bibfield  {journal} {\bibinfo  {journal} {Phys. Rev. D}\ }\textbf {\bibinfo {volume} {102}},\ \bibinfo {pages} {066004} (\bibinfo {year} {2020})}\BibitemShut {NoStop}%
\bibitem [{\citenamefont {Skenderis}\ and\ \citenamefont {Taylor}(2008)}]{Skenderis_2008}%
  \BibitemOpen
  \bibfield  {author} {\bibinfo {author} {\bibfnamefont {K.}~\bibnamefont {Skenderis}}\ and\ \bibinfo {author} {\bibfnamefont {M.}~\bibnamefont {Taylor}},\ }\bibfield  {title} {\bibinfo {title} {The fuzzball proposal for black holes},\ }\href {https://doi.org/10.1016/j.physrep.2008.08.001} {\bibfield  {journal} {\bibinfo  {journal} {Physics Reports}\ }\textbf {\bibinfo {volume} {467}},\ \bibinfo {pages} {117–171} (\bibinfo {year} {2008})}\BibitemShut {NoStop}%
\bibitem [{\citenamefont {Raju}(2021)}]{raju2021lessonsinformationparadox}%
  \BibitemOpen
  \bibfield  {author} {\bibinfo {author} {\bibfnamefont {S.}~\bibnamefont {Raju}},\ }\href {https://arxiv.org/abs/2012.05770} {\bibinfo {title} {Lessons from the information paradox}} (\bibinfo {year} {2021}),\ \Eprint {https://arxiv.org/abs/2012.05770} {arXiv:2012.05770 [hep-th]} \BibitemShut {NoStop}%
\bibitem [{\citenamefont {Geng}\ \emph {et~al.}(2021)\citenamefont {Geng}, \citenamefont {Nomura},\ and\ \citenamefont {Sun}}]{Geng_2021}%
  \BibitemOpen
  \bibfield  {author} {\bibinfo {author} {\bibfnamefont {H.}~\bibnamefont {Geng}}, \bibinfo {author} {\bibfnamefont {Y.}~\bibnamefont {Nomura}},\ and\ \bibinfo {author} {\bibfnamefont {H.-Y.}\ \bibnamefont {Sun}},\ }\bibfield  {title} {\bibinfo {title} {Information paradox and its resolution in de sitter holography},\ }\bibfield  {journal} {\bibinfo  {journal} {Physical Review D}\ }\textbf {\bibinfo {volume} {103}},\ \href {https://doi.org/10.1103/physrevd.103.126004} {10.1103/physrevd.103.126004} (\bibinfo {year} {2021})\BibitemShut {NoStop}%
\bibitem [{\citenamefont {Miller}(2018)}]{miller_2018_black}%
  \BibitemOpen
  \bibfield  {author} {\bibinfo {author} {\bibfnamefont {N.}~\bibnamefont {Miller}},\ }\href {https://scholar.harvard.edu/files/noahmiller/files/firewall.pdf} {\bibinfo {title} {Black holes, hawking radiation, and the firewall}} (\bibinfo {year} {2018})\BibitemShut {NoStop}%
\end{thebibliography}%

\end{document}